# An Adversarial Robustness Benchmark for Enterprise Network Intrusion Detection

João Vitorino[0000-0002-4968-3653], Miguel Silva[0009-0008-6630-9939], Eva Maia[0000-0002-8075-531X] and Isabel Praça[0000-0002-2519-9859]

Research Group on Intelligent Engineering and Computing for Advanced Innovation and Development (GECAD), School of Engineering, Polytechnic of Porto (ISEP/IPP), 4249-015 Porto, Portugal
{jpmvo,mdgsa,egm,icp}@isep.ipp.pt

**Abstract.** As cyber-attacks become more sophisticated, improving the robustness of Machine Learning (ML) models must be a priority for enterprises of all sizes. To reliably compare the robustness of different ML models for cyber-attack detection in enterprise computer networks, they must be evaluated in standardized conditions. This work presents a methodical adversarial robustness benchmark of multiple decision tree ensembles with constrained adversarial examples generated from standard datasets. The robustness of regularly and adversarially trained RF, XGB, LGBM, and EBM models was evaluated on the original CICIDS2017 dataset, a corrected version of it designated as NewCICIDS, and the HIKARI dataset, which contains more recent network traffic. NewCICIDS led to models with a better performance, especially XGB and EBM, but RF and LGBM were less robust against the more recent cyber-attacks of HIKARI. Overall, the robustness of the models to adversarial cyber-attack examples was improved without their generalization to regular traffic being affected, enabling a reliable detection of suspicious activity without costly increases of false alarms.

**Keywords:** machine learning, enterprise networks, adversarial attacks, adversarial training, cybersecurity

## 1 Introduction

Protecting digital assets and business processes is a priority for enterprises of all sizes. As cyber-attacks become more sophisticated, network intrusion detection systems stand out as a critical security component to monitor network traffic and identify suspicious activity [1]. Artificial Intelligence (AI), and more specifically Machine Learning (ML), has become significantly valuable to strengthen enterprise network security. ML models can be used to tackle the growing number of threats by performing anomaly detection and even classifying the cyber-attacks targeting an enterprise [2].

In the network intrusion detection domain, ML models based on ensembles of decision trees are very well-established. Bagging ensembles like Random Forest (RF) and boosting ensembles like Extreme Gradient Boosting (XGB) are reliable and computationally efficient models that are commonly used to detect and classify cyber-attacks in enterprise-scale computer networks [3], [4].

However, an attacker may craft an adversarial cyber-attack example with specialized inputs capable of evading detection and disrupting the confidentiality, integrity, and availability of the data and business processes of an enterprise. Adversarial ML is a rapidly growing research field that addresses these disruptions by studying the adversarial attacks that attempt to exploit the vulnerabilities of ML models and the possible defense strategies to improve robustness against such attacks [5], [6].

Even though there are several studies that use standard benchmark datasets to compare the performance of ML models for network intrusion detection, there is still a lack of consistency in the studies that analyze their robustness [7], [8]. Different studies follow distinct approaches to evaluate robustness, preventing researchers from knowing which models are the most suitable for their specific computer networks, and which adversarial attacks could be used against them in a real scenario [9], [10]. Therefore, to reliably compare the robustness of different types of ML models with different datasets, they must be evaluated in standardized conditions.

This work presents a methodical adversarial robustness benchmark of multiple decision tree ensembles with constrained adversarial examples generated from standard datasets for network intrusion detection. The robustness of regularly and adversarially trained RF, XGB, Light Gradient Boosting Machine (LGBM), and Explainable Boosting Machine (EBM) models is evaluated on the original CICIDS2017 dataset, a corrected version of it designated as NewCICIDS, and the HIKARI dataset, which contains more recent network traffic. The results obtained in each dataset are compared to analyze their suitability for network intrusion detection tasks, and the robustness of the models to adversarial cyber-attack examples is assessed, to verify if it can be improved without their generalization to regular network traffic being affected.

The present paper is organized into multiple sections, meant to enable researchers to replicate this benchmark and perform trustworthy comparisons with the results of future studies. Section 2 provides a survey of previous work on the use of ML for enterprise network intrusion detection and the standard datasets. Section 3 describes the data pre-processing and selected features, the benchmark methodology, and the fine-tuning of the models. Section 4 presents an analysis of the obtained results for each dataset. Finally, Section 5 addresses the main conclusions and future research topics.

## 2    Related Work

To perform a reliable benchmark of RF, XGB, LGBM, and EBM, it is important to understand the results and conclusions of previous work on the use of ML for network intrusion detection in enterprise computer networks.

Recent studies [11]–[13] mostly evaluate ML models using the publicly available CICIDS2017 dataset, which is very well-established across the scientific community. Despite not being very recent, it continues to be used to compare the performance of novel models with the state-of-the-art results of previous models of the previous years. Since this dataset is so widely used, researchers have also carefully analyzed it and performed corrections to some of the network traffic flows it contains, publishing a corrected version designated as NewCICIDS [14], [15]. Despite the limited use of this newer version, it was used in [16] to evaluate the robustness of decision tree ensembles when confronted with adversarial attacks. The results exhibited a lack of robustness in these ensembles, so it is essential to further improve their training processes.

As new cyber-attacks and adversarial methods are encountered, it is essential to train ML models with the most up-to-date datasets containing high-quality data recordings [17]. To develop more robust cybersecurity solutions, recent studies start to use other datasets like HIKARI because there are new types of attacks that are starting to be used against modern enterprises [18], [19].

In a recent study addressing botnet detection [20], the researchers trained lightweight models using both the HIKARI dataset and the CTU-13 dataset. The authors experimented dimensionality reduction techniques, and the models reached an accuracy of 99% with all the features and 96% with a reduced number of features. Expanding to other types of cyber-attacks, in [21], the CICIDS2017 and HIKARI datasets were combined to improve data quality. This study compared various models, including RF, XGB, Logistic Regression, Deep Neural Network, and Long Short-Term Memory, and the best precision was 99% and the accuracy 86%. These results are substantially lower than in the previous study that only used the HIKARI dataset, so they may be caused by the increased diversity of cyber-attacks of the combined datasets.

To choose the best models capable of dealing with the unbalanced data of these datasets, in [22], the HIKARI dataset was used for training RF, XGB, Multilayer Perceptron, and Convolutional Neural Network. The authors concluded that these models may not effectively detect zero-day attacks, so it is necessary to improve their training processes to make them more robust. Additionally, in [23], the HIKARI dataset was also used to develop classifiers capable of detecting out-of-distribution data, which presents similarities to zero-day attacks. Several tree ensembles, such as RF and Gradient Boosted Decision Trees (GDBT), were evaluated. The authors' version of GBDT reached the highest accuracy, 99.57%, and area under the ROC curve, 90.30%. On the other hand, RF achieved the higher area under the precision and recall curve, 83.35%, denoting that, in some evaluation metrics, the lightweight ensembles can reach better results than the more complex ensembles that perform gradient boosting.

Overall, the recent studies on network intrusion have demonstrated that using tree ensembles is a promising approach to detect suspicious activity in enterprise computer networks. The commonly used models include stacking and bagging ensembles, as well as gradient boosting ensembles, which can have very good results against regular network traffic flows [19]. However, to the best of our knowledge, no previous work has analyzed how the time-related characteristics of the three considered datasets affect the robustness of RF, XGB, LGBM, and EBM against perturbed network traffic flows.

## 3 Methods

This section describes the considered datasets, the data preprocessing stage, and the utilized models and adversarial method. The work was carried out on a common machine with 16GB of RAM, a 6-core CPU and a 4GB GPU. The implementation relied on the Python programming language and the following libraries: *numpy* and *pandas* for general data manipulation, *scikit-learn* for the implementation of RF, *xgboost* for XGB, *lightgbm* for LGBM, and *interpret* for EBM.

### 3.1   Datasets and Data Preprocessing

Due to their use across several studies, three standard datasets for binary network traffic classification were considered for the benchmark: CICIDS2017, NewCICIDS, and HIKARI. The main characteristics of these datasets are briefly described below.

CICIDS2017 [24] is a very highly used dataset that contains common cyber-attacks performed in an enterprise computer network. It includes multiple captures of benign activity and several types of probing, brute-force, and DoS attacks, which were recorded in 2017 in an heterogenous testbed environment with 12 interacting machines. The network traffic flows were converted to a tabular data format using the CICFlowMeter [25] tool, provided by the Canadian Institute for Cybersecurity. This process resulted in 872105 data samples of the benign class and 266507 of the malicious class, in the combined dataset of the Tuesday and Wednesday captures.

Even though CICIDS2017 continues to be used as a standard benchmark dataset to compare the performance of novel ML models with baseline models from previous studies, some discrepancies have been noticed on a portion of the attack vectors it contains. A corrected version of this dataset has been created to address this issue and provide more realistic network traffic flows, being designated as NewCICIDS [14], [15]. It includes the same types of cyber-attacks as the original dataset, but it has a reduced size, with 638432 benign samples and 106538 malicious samples.

A more recent dataset, HIKARI [26], is starting to be used in various studies because it includes cyber-attacks that have started to be performed in more recent years. It contains probing and brute-force attacks, as well as benign background traffic of the normal operation of an enterprise network that uses the HTTPS communication protocol to encrypt network traffic. The data was recorded in 2021 to tackle the lack of datasets containing application-layer attacks on encrypted traffic, using similar features to those utilized in CICIDS2017 and NewCICIDS. The resulting network traffic flows correspond to 214904 benign samples and 13349 malicious samples, so HIKARI has a higher data imbalance than the previous two datasets, representing more realistic conditions for enterprise-scale network intrusion detection.

Before the three datasets could be used, a data preprocessing phase was required to create stratified training and holdout sets with 70% and 30% of the data, respectively. In addition to removing rows with missing data, it was necessary to select relevant and unbiased features. A study [27] has analyzed the feature importance rankings of the more than 80 features of the HIKARI dataset, observing that the most impactful ones represented time-related characteristics. By only using these features, the training and inference time of multiple ML models were greatly reduced without a significant decrease in their performance in a holdout evaluation. Therefore, it is possible to select only these features and still achieve a good generalization.

Table 1 provides an overview of the selected time-related characteristics of network traffic flows. From 7 main characteristics, 24 features were selected, based on the feature importance rankings of the considered study. In the three datasets, the forward part of a flow corresponds to a client machine that opens a connection with the server, sending network packets. Likewise, the backward part corresponds to the packets sent by the server back to the client within that connection. The full connection will be classified as either a benign flow that is part of the normal operation of the network or a malicious flow in which the client sent ill-intentioned packets. Regarding the IAT keyword, it corresponds to the Inter-Arrival Time, the elapsed time between the arrival of two subsequent network packets within a flow.

Table 1. Main characteristics of the utilized datasets.

| Characteristic | Description | Selected Features | | | | |
|---|---|---|---|---|---|---|
| | | Total | Mean | Std | Max | Min |
| Flow Packet IAT | Packet IAT of the full connection | No | Yes | Yes | Yes | Yes |
| Forward Packet IAT | Packet IAT of the client | Yes | Yes | Yes | Yes | Yes |
| Backward Packet IAT | Packet IAT of the server | Yes | Yes | Yes | Yes | Yes |
| Forward Bulk Rate | Transmission rate of the client | No | Yes | No | No | No |
| Backward Bulk Rate | Transmission rate of the server | No | Yes | No | No | No |
| Flow Active Time | Transmission time of the full connection | No | Yes | Yes | Yes | Yes |
| Flow Idle Time | Inactive time of the full connection | No | Yes | Yes | Yes | Yes |

### 3.2 Benchmark Methodology

The robustness analysis methodology introduced in [28] was followed to ensure an unbiased benchmark of the considered ML models. It includes both a regular training process and an adversarial training process, which is a well-established adversarial defense strategy. In the former, the original training set of a certain dataset is used to train, fine-tune, and validate an ML model. In the latter, data augmentation is performed by creating simple perturbations in the original training set, resulting in an adversarial training set that contains both original data samples and slightly perturbed data samples.

Afterwards, the considered methodology establishes a performance evaluation in both normal conditions and during a direct attack to the models. In the former, the models perform predictions of the data samples in the regular holdout set of a certain dataset, and several standard evaluation metrics are computed. In the latter, a full adversarial evasion attack is performed against each model, with specialized perturbations to deceive that specific model. Since different models are susceptible to different perturbations, the attacks result in model-specific adversarial holdout sets. In the case of network intrusion detection, these attacks are targeted, attempting to cause misclassifications from the malicious class to the target benign class.

The adversarial examples were generated using the Adaptative Perturbation Pattern Method (A2PM) [29]. It relies on pattern sequences that learn the characteristics of each class and create constrained data perturbations, according to the provided information about the feature set, which corresponds to a gray-box setting. The patterns record the value intervals of different feature subsets, which are then used to ensure that the perturbations take the correlations of the features into account, generating realistic adversarial examples. Therefore, when applied to network intrusion detection, the patterns iteratively optimize the perturbations that are performed on each feature of a network traffic flow according to the constraints of a computer network.

For adversarial training, a simple function provided by the method was used to create a perturbation in each malicious sample of a training set, performing data augmentation. Hence, a model was able to learn not only from a sample, but also from a simple variation of it. Fig. 1 provides an overview of the creation of an adversarial training set that is common to all models. Starting from a regular training set with 70% of a dataset, another set of the same size can be obtained, with a perturbation in each sample.

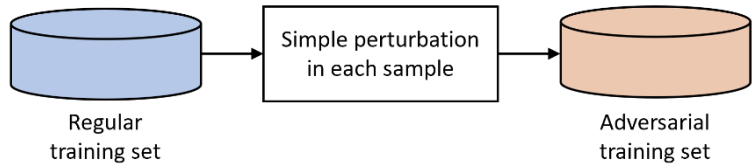

**Fig. 1.** Creation of a simple adversarial training set.

To perform adversarial evasion attacks specific to each model, the full A2PM attack created as many data perturbations as necessary in a holdout set until every malicious sample was misclassified as benign or a total of 15 attack iterations were performed. No more iterations were allowed because a high number of requests to a specific server would increase the risk of the anomalous behavior being noticed by the security practitioners overseeing the networking infrastructure of an enterprise network [10]. Fig. 2 provides an overview of the creation of model-specific adversarial holdout sets that are used for the benchmark. Starting from a regular holdout set with 30% of a dataset, several other sets of the same size can be obtained, with many specialized data perturbations to cause misclassifications in a certain ML model.

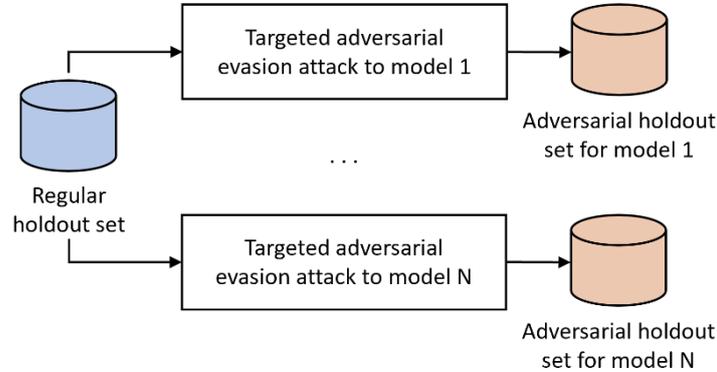

**Fig. 2.** Creation of model-specific adversarial holdout sets.

### 3.3 Models and Fine-tuning

Due to their well-established performance in enterprise network intrusion detection and the good results obtained in the surveyed studies, four supervised tree ensembles were considered for the benchmark: RF, XGB, LGBM, and EBM.

The optimal configuration for each model and each dataset were obtained via a grid search with well-established hyperparameter combinations for binary network traffic classification, and the best hyperparameters were determined through a 5-fold cross-validation. Five stratified subsets were created, each with 20% of a training set. Then, five distinct iterations were performed, each training a model with four subsets and validating it with the remaining one. Due to its adequacy for unbalanced data and consolidation of precision and recall, the F1-score was selected as the validation metric. After the fine-tuning process, each model was retrained with a complete training set, so it was ready for the benchmark with the regular and adversarial holdout sets, which also included other evaluation metrics, such as accuracy and false positive rate.

**Random Forest.** RF [30] is a supervised ensemble created through bagging and using the Gini Impurity criterion to calculate the best node splits. Each individual tree performs a prediction according to a feature subset, and the most common vote is chosen. RF is based on the concept that the collective decisions of many trees will be better than the decisions of just one. Table 2 summarizes the fine-tuned configuration.

Table 2. Summary of RF configuration.

| Parameter | Value |
|---|---|
| Criterion | Gini Impurity |
| No. of estimators | 100 |
| Max. features | 4 |
| Max. depth of a tree | 8 to 16 |
| Min. samples in a leaf | 2 |

**Extreme Gradient Boosting.** XGB [31] performs gradient boosting using a supervised ensemble with a level-wise growth strategy. The nodes within each tree are split level by level, using the Histogram method to compute fast histogram-based approximations and seeking to minimize the Cross-Entropy loss function during its training. Table 3 summarizes the fine-tuned configuration.

Table 3. Summary of XGB configuration.

| Parameter | Value |
| --- | --- |
| Method | Histogram |
| Loss function | Cross-Entropy |
| No. of estimators | 100 |
| Max. depth of a tree | 4 to 16 |
| Min. leaf weight | 1 |
| Min. loss reduction | 0.01 |
| Learning rate | 0.1 to 0.3 |
| Feature subsample | 0.7 to 0.8 |

**Light Gradient Boosting Machine.** LGBM [32] also uses a supervised ensemble to perform gradient boosting. The nodes are split using a leaf-wise strategy for a best-first approach, performing the split with the higher loss reduction. LGBM uses Gradient-based One-Side Sampling (GOSS) to build the decision trees, which is computationally lighter than previous methods and therefore provides a faster training process. Table 4 summarizes the fine-tuned configuration.

Table 4. Summary of LGBM configuration.

| Parameter | Value |
| --- | --- |
| Method | GOSS |
| Loss function | Cross-Entropy |
| No. of estimators | 100 |
| Max. leaves in a tree | 15 |
| Min. samples in a leaf | 20 |
| Min. loss reduction | 0.01 |
| Learning rate | 0.1 to 0.2 |
| Feature subsample | 0.7 to 0.8 |

**Explainable Boosting Machine.** EBM [33] is a generalized additive model that performs cyclic gradient boosting with a tree ensemble. Unlike the other three black-box models, EBM is a glass-box model that remains explainable and interpretable during the inference phase [34]. Each feature contributes to a prediction in an additive manner that enables their individual contribution to be measured and explained. Table 5 summarizes the fine-tuned configuration.

Table 5. Summary of EBM configuration.

| Parameter | Value |
|---|---|
| Loss function | Cross-Entropy |
| No. of estimators | 100 |
| Max. number of bins | 256 |
| Max. leaves in a tree | 7 to 15 |
| Min. samples in a leaf | 2 |
| Learning rate | 0.1 |

## 4 Results and Discussion

This section presents and discusses the results obtained by evaluating the performance of the ML models created through regular and adversarial training. The evaluation considers the regular holdout set of the CICIDS2017, NewCICIDS, and HIKARI datasets, as well as the model-specific adversarial holdout sets.

### 4.1 CICIDS2017

The models trained with the CICIDS2017 dataset obtained very high results across several standard evaluation metrics. Despite only using 24 time-related features of the dataset, they enabled all four models to detect the anomalous behavior of most malicious flows, distinguishing cyber-attacks from benign activity and reaching F1-scores over 89%. Nonetheless, when adversarial attacks were performed against these models, their precision and recall exhibited significant declines that resulted in F1-scores lower than 0.1% after the attack iterations were complete. This failure to detect adversarial examples suggests that tree ensembles are inherently vulnerable to modifications of the time-related characteristics of network traffic flows.

On the other hand, the models created through adversarial training had substantially lower declines, preserving their precision above 97%. Even though the recall of EBM was only approximately 60% when attacked, RF, XGB, and LGBM all retained a higher recall above 73%. Hence, by training with a simple perturbation per malicious sample, the robustness of the models was improved, and most malicious flows could not evade detection. Regarding benign flows, it is important to note that the false positive rates were decreased to below 0.40%, which indicates that deploying these models in a real computer network could lead to less false alarms and therefore less unnecessary mitigation measures that would be costly for an enterprise.

Table 6 provides the obtained results for the models trained with the CICIDS2017 dataset, considering standard evaluation metrics for binary network traffic classification. The ACC, PRC, RCL, F1S, and FPR columns correspond to accuracy, precision, recall, F1-score, and false positive rate. The optimal result would be 100% for all metrics except the false positive rate, which should be as close to 0% as possible. Additionally, the results achieved by the adversarially trained models on the adversarial holdout sets are highlighted in bold.

Table 6. Obtained results for the CICIDS2017 dataset.

| Model | Training | Attacked | Evaluation Metrics (%) | | | | |
|---|---|---|---|---|---|---|---|
| | | | ACC | PRC | RCL | F1S | FPR |
| RF | Regular | No | 95.21 | 90.74 | 88.59 | 89.65 | 2.76 |
| | | Yes | 74.48 | 0.03 | 0.01 | 0.01 | 2.76 |
| | Adversarial | No | 93.84 | 99.20 | 74.30 | 84.96 | 0.18 |
| | | Yes | **93.80** | **99.19** | **74.12** | **84.84** | **0.18** |
| XGB | Regular | No | 95.29 | 90.73 | 88.94 | 89.83 | 2.78 |
| | | Yes | 74.48 | 0.36 | 0.03 | 0.06 | 2.78 |
| | Adversarial | No | 94.67 | 98.93 | 78.07 | 87.27 | 0.26 |
| | | Yes | **94.30** | **98.91** | **76.5** | **86.28** | **0.26** |
| LGBM | Regular | No | 94.95 | 90.12 | 88.09 | 89.09 | 2.95 |
| | | Yes | 74.35 | 0.57 | 0.05 | 0.09 | 2.95 |
| | Adversarial | No | 94.19 | 98.93 | 76.01 | 85.97 | 0.25 |
| | | Yes | **93.50** | **98.89** | **73.05** | **84.03** | **0.25** |
| EBM | Regular | No | 94.98 | 90.43 | 87.86 | 89.13 | 2.84 |
| | | Yes | 74.42 | 0.08 | 0.01 | 0.01 | 2.84 |
| | Adversarial | No | 94.4 | 98.43 | 77.31 | 86.6 | 0.38 |
| | | Yes | **90.13** | **97.96** | **60.08** | **74.48** | **0.38** |

### 4.2 NewCICIDS

The models trained with the corrected version of CICIDS2017 exhibited much better results than those of the original dataset. Training with the corrected network traffic flows of NewCICIDS led all four models to achieve F1-scores higher than 99% on the regular holdout set, and their false positive rates did not exceed 0.20%. Despite the performance of the models being significantly decreased in the model-specific adversarial holdout sets, it was still slightly better than the decline observed in the original dataset. It is pertinent to highlight the better robustness of the regularly trained EBM, which retained a precision of over 31% throughout the adversarial evasion attack just by training with the corrected flows of NewCICIDS.

As before, performing adversarial training led to a great improvement in the robustness of the models. This defense strategy enabled the detection of most adversarial cyber-attack examples, reducing the number of misclassifications that would be harmful for an enterprise. Even though the recall of the adversarially trained RF and XGB was slightly decreased, they preserved their precision of 99.88% and 99.90%, without this metric being decreased by the attack. Since only very few of the perturbed malicious flows were misclassified, the functionality of those cyber-attacks would be prevented in a real enterprise communication network. Furthermore, RF and XGB achieved the best false positive rates, 0.06% and 0.05%, respectively, which indicates that the corrected dataset also led to models with better generalization.

Table 7 provides the obtained results for the models trained with the NewCICIDS dataset, highlighting the precision of the regularly trained EBM and the results of all the adversarially trained models.

Table 7. Obtained results for the NewCICIDS dataset.

| Model | Training | Attacked | Evaluation Metrics (%) | | | | |
|---|---|---|---|---|---|---|---|
| | | | ACC | PRC | RCL | F1S | FPR |
| RF | Regular | No | 99.90 | 99.81 | 99.92 | 99.87 | 0.11 |
| | | Yes | 64.19 | 0.01 | 0.01 | 0.01 | 0.11 |
| | Adversarial | No | 99.67 | 99.88 | 99.20 | 99.54 | 0.06 |
| | | Yes | **99.63** | **99.88** | **99.08** | **99.48** | **0.06** |
| XGB | Regular | No | 99.94 | 99.89 | 99.94 | 99.92 | 0.06 |
| | | Yes | 64.22 | 2.59 | 0.01 | 0.01 | 0.06 |
| | Adversarial | No | 99.93 | 99.90 | 99.89 | 99.90 | 0.05 |
| | | Yes | **99.84** | **99.90** | **99.64** | **99.77** | **0.05** |
| LGBM | Regular | No | 99.79 | 99.63 | 99.79 | 99.71 | 0.20 |
| | | Yes | 64.14 | 7.36 | 0.03 | 0.06 | 0.20 |
| | Adversarial | No | 99.72 | 99.67 | 99.54 | 99.60 | 0.19 |
| | | Yes | **94.91** | **99.61** | **86.10** | **92.36** | **0.19** |
| EBM | Regular | No | 99.86 | 99.76 | 99.84 | 99.80 | 0.14 |
| | | Yes | 64.21 | **31.48** | 0.11 | 0.22 | 0.14 |
| | Adversarial | No | 99.80 | 99.74 | 99.71 | 99.72 | 0.15 |
| | | Yes | **98.13** | **99.72** | **95.02** | **97.32** | **0.15** |

### 4.3 HIKARI

The models trained with the more recent HIKARI dataset obtained F1-scores between 83% and 84%, which are lower than those of the previous datasets, but are still reasonably high for a binary network traffic classification task. The targeted adversarial evasion attack caused the recall and precision of all four models to decrease, but XGB was able to retain a precision of over 58% and EBM of over 98%. Since their false positive rates were near 0.01%, the results denote that more than half of the adversarial examples were detected and there were very benign flows mistakenly predicted as malicious, which is very important for an enterprise-scale computer network.

Despite also having equivalent false positive rates, the adversarially trained models did not exhibit high increases in their robustness. When attacked, the F1-scores of RF, XGB, LGBM, and EBM, were approximately 82%, 62%, 27%, and 63%, respectively. These results are substantially lower than those obtained in the previous datasets, suggesting that the greater complexity of the more recent cyber-attacks makes it more difficult to distinguish them from benign flows that are part of the normal operation of an enterprise network. Therefore, adversarial training is not always guaranteed to help ML models achieve an adversarially robust generalization. It is pertinent to carefully evaluate their performance, assessing if they exhibit a good generalization to regular traffic and a good robustness to adversarially perturbed traffic.

Table 8 provides the obtained results for the models trained with the HIKARI dataset, highlighting the precision of the regularly trained XGB and EBM and the results of all the adversarially trained models.

Table 8. Obtained results for the HIKARI dataset.

| Model | Training | Attacked | Evaluation Metrics (%) | | | | |
|---|---|---|---|---|---|---|---|
| | | | ACC | PRC | RCL | F1S | FPR |
| RF | Regular | No | 98.34 | 99.79 | 71.84 | 83.54 | 0.01 |
| | | Yes | 94.14 | 0.01 | 0.01 | 0.01 | 0.01 |
| | Adversarial | No | 98.33 | 99.90 | 71.59 | 83.40 | 0.01 |
| | | Yes | **98.21** | **99.89** | **69.46** | **81.94** | **0.01** |
| XGB | Regular | No | 98.35 | 99.83 | 71.91 | 83.60 | 0.01 |
| | | Yes | 94.15 | **58.33** | 0.17 | 0.35 | 0.01 |
| | Adversarial | No | 98.36 | 99.86 | 72.01 | 83.68 | 0.01 |
| | | Yes | **96.76** | **99.78** | **44.72** | **61.76** | **0.01** |
| LGBM | Regular | No | 98.36 | 99.86 | 72.13 | 83.76 | 0.01 |
| | | Yes | 94.15 | 0.01 | 0.01 | 0.01 | 0.01 |
| | Adversarial | No | 98.35 | 99.72 | 72.01 | 83.63 | 0.01 |
| | | Yes | **95.05** | **98.74** | **15.63** | **26.99** | **0.01** |
| EBM | Regular | No | 98.35 | 99.76 | 72.01 | 83.64 | 0.01 |
| | | Yes | 95.01 | **98.84** | 14.86 | 25.83 | 0.01 |
| | Adversarial | No | 98.35 | 99.69 | 71.94 | 83.57 | 0.01 |
| | | Yes | **96.84** | **99.52** | **46.14** | **63.05** | **0.01** |

## 5   Conclusions

This work benchmarked the robustness of multiple decision tree ensembles for enterprise network intrusion detection. Regularly and adversarially trained RF, XGB, LGBM, and EBM models were fine-tuned and evaluated on the original CICIDS2017 dataset, the corrected NewCICIDS dataset, and the HIKARI dataset with more recent network traffic. Targeted adversarial evasion attacks were performed using A2PM, and the results obtained in the adversarial holdout sets were compared to those of the regular holdout sets, assessing if the models correctly classified perturbed data samples and preserved high evaluation metrics, indicating a good robustness.

The best results across several evaluation metrics were achieved in the corrected version of CICIDS2017, which provided substantial improvements in both regular samples and perturbed samples, in comparison with the original dataset. Even though the recall of the adversarially trained RF and XGB was slightly decreased, these models were able to preserve best precision, 99.88% and 99.90%. Hence, the adversarially trained models were able to detect most of the perturbed malicious flows, which would prevent the functionality of those cyber-attacks in a real enterprise network.

However, when facing the more recent network traffic of the HIKARI dataset, the ML models were less robust. XGB and EBM preserved a reasonably good precision when attacked, but RF and LGBM exhibited numerous misclassifications. Despite the best false positive rates being achieved in HIKARI, these worse results suggest that the greater complexity of the more recent cyber-attacks makes it more difficult to distinguish them from benign flows. Therefore, in addition to an adversarial training process, other adversarial defense strategies may be needed.

Overall, the robustness of the four ML models to adversarial cyber-attack examples was improved without their generalization to regular traffic being affected, enabling a reliable detection of suspicious activity in enterprise networks without costly increases of false alarms. In the future, it could be valuable to explore the intrinsic explainability capabilities of EBM and apply ad hoc explainability methods to RF, XGB, and LGBM, enabling a better understanding of the reasoning behind their misclassifications. To further contribute to adversarial ML research, it is important to also benchmark the adversarial robustness of these tree ensembles for multi-class classification and compare them with other types of ML models, including deep learning models.

**Acknowledgments.** This work has been supported by the UIDB/00760/2020 and UIDP/00760/2020 projects.

```
```